\documentclass[final,3p,times,twocolumn]{elsarticle}
\usepackage{graphicx}
\usepackage{amssymb}
\usepackage{amsmath}
\usepackage{epsfig}
\usepackage{longtable}
\usepackage{rotating}
\usepackage{lscape}
\usepackage{booktabs}
\journal{Physics Letters B}
\begin{document}
\begin{frontmatter}



\title{Nuclear state densities of odd-mass heavy nuclei in the shell model Monte Carlo approach}


\author[khas,yale]{C. \"{O}zen\corref{cor1}}
%\ead{cem.ozen@khas.edu.tr}
\author[yale]{Y. Alhassid}
\author[chiba]{H.~Nakada}
%\ead{yoram.alhassid@yale.edu}


\address[khas]{Faculty of Engineering and Natural Sciences, Kadir Has University, Istanbul 34083, Turkey}
\address[yale]{Center for Theoretical Physics, Sloane Physics Laboratory, Yale University,\\ New Haven, CT 06520, USA}
\address[chiba]{Department of Physics, Graduate School of Science, Chiba University, Inage, Chiba, 263-8522, Japan}

\begin{abstract}
%The projection on an odd number of particles in the shell model Monte Carlo approach leads to a sign problem at low temperatures, making it impractical to extract the ground-state energy in direct Monte Carlo calculations, and thus prohibiting the calculation of state densities of odd-even heavy nuclei.
While the shell model Monte Carlo approach has been successful in the microscopic calculation of nuclear state densities, it has been difficult to calculate accurately state densities of odd-even heavy nuclei. This is because the projection on an odd number of particles in the shell model Monte Carlo method leads to a sign problem at low temperatures, making it impractical to extract the ground-state energy in direct Monte Carlo calculations. We show that the ground-state energy can be extracted to a good precision by using level counting data at low excitation energies and the neutron resonance data at the neutron threshold energy. This allows us to extend recent applications of the shell-model Monte Carlo method in even-even rare-earth nuclei to the odd-even isotopic chains of $^{149-155}$Sm and $^{143-149}$Nd. We calculate the state densities of the odd-even samarium and neodymium isotopes and find close agreement with the state densities extracted from experimental data.
\end{abstract}

\begin{keyword}
nuclear shell model, quantum Monte Carlo, level density, back-shifted Bethe formula

\end{keyword}
\end{frontmatter}
% \linenumbers


%\maketitle
%
\section{Introduction}
\label{intro}

Reliable microscopic calculation of nuclear state densities of heavy nuclei is a challenging task because it often requires the inclusion of correlations beyond the mean-field approximation. Correlation effects can be taken into account in the context of the configuration-interaction (CI) shell model approach, but the size of the required model space in heavy nuclei is prohibitively large for direct diagonalization of the CI shell model Hamiltonian. This limitation can be overcome in part by using the shell model Monte Carlo (SMMC) method~\cite{JOHN92,LANG93,ALH94,KOO98,ALH01}. The SMMC method enables the calculations of statistical nuclear properties, and in particular of level densities in very large model spaces~\cite{NAK97,ORM97,ALH99,ALH07,OZE07}.

Fermionic Monte Carlo methods are often hampered by the so-called sign problem, which leads to large and uncontrollable fluctuations of observables at low temperatures~\cite{JOHN92,LANG93,WHI89,LIN92}. Statistical and collective properties of nuclei can be reliably calculated~\cite{ALH96,NAK97} by employing a class of interactions that have a good Monte Carlo sign in the grand-canonical formulation~\cite{LANG93,ALH96}. The projection on an even number of particles keeps the good sign of the interaction, allowing accurate calculations for even-even nuclei.  However, the projection on an odd number of particles leads to a new sign problem, making it difficult to calculate thermal observables at low temperatures for odd-even and odd-odd nuclei. In particular, the ground-state energy of the odd-particle system cannot be extracted from direct SMMC calculations.  A method to extract the ground-state energy of the odd-particle system from the imaginary-time Green's functions of the even-particle system was recently introduced in Ref.~\cite{MUK12} and applied successfully to medium-mass nuclei. However, the application of this method to heavy nuclei is computationally intensive and requires additional development. In this work, we describe a practical method that enables us to determine the ground-state energy using a complete set of experimentally known low-lying levels and the neutron resonance data at the neutron threshold energy. We then extend the recent SMMC state density calculations of even-even nuclei in the rare-earth region~\cite{ALH08,OZE12} to include odd-even isotope chains of $^{149-155}$Sm and $^{143-149}$Nd. We find close agreement with the state densities that are extracted from experimental data.

\section{Choice of Model Space and Interaction}
%\label{sec:1}
In rare-earth nuclei we use the model space of Refs.~\cite{ALH08,OZE12} spanned by the single-particle orbitals $0g_{7/2}$, $1d_{5/2}$, $1d_{3/2}$, $2s_{1/2}$, $0h_{11/2}$ and $1f_{7/2}$ for protons, and the orbitals $0h_{11/2}$, $0h_{9/2}$, $1f_{7/2}$, $1f_{5/2}$,
$2p_{3/2}$, $2p_{1/2}$, $0i_{13/2}$, and $1g_{9/2}$ for neutrons. We have chosen these orbitals by the requirement that their occupation probabilities in well-deformed nuclei be between 0.9 and 0.1. The effect of other orbitals is accounted for by the renormalization of the interaction. The bare single-particle energies in the shell-model Hamiltonian are obtained so as to reproduce the single-particle energies in a spherical Woods-Saxon plus spin-orbit potential in the spherical Hartree-Fock approximation.

The effective interaction consists of monopole pairing and multipole-multipole terms~\cite{ALH08}
\begin{equation}
 - \!\! \sum_{\nu=p,n} g_\nu P^\dagger_\nu P_\nu
 - \!\! \sum_\lambda \chi_\lambda : (O_{\lambda;p} + O_{\lambda;n})\cdot
 (O_{\lambda;p} + O_{\lambda;n})\!: \;,
\label{Eq:Ham}
\end{equation}
where the pair creation operator $P^\dagger_\nu$ and the multipole operator $O_{\lambda;\nu}$ are given by
\begin{subequations}
\begin{equation}
P^\dagger_\nu = \sum_{nljm}(-)^{j+m+l}a^\dagger_{\alpha jm;\nu}
a^\dagger_{\alpha j-m;\nu}\;,
\end{equation}
\begin{equation}
O_{\lambda;\nu}=\frac{1}{\sqrt{2\lambda+1}}
\sum_{ab}\langle j_a||\frac{dV_\mathrm{WS}}{dr}Y_\lambda||j_b\rangle
[a^\dagger_{\alpha j_a;\nu}\times \tilde{a}_{\alpha j_b;\nu}]^{(\lambda)}
\end{equation}
\end{subequations}
with $\tilde{a}_{jm}=(-)^{j+m}a_{j-m}$. In Eq.~(\ref{Eq:Ham}), $:\,:$ denotes normal ordering and $V_\mathrm{WS}$ represents the Woods-Saxon potential.  The pairing strengths is expressed as  $g_\nu= \gamma \bar{g}_\nu$,
where $\gamma$ is a renormalization factor and $\bar{g}_p=10.9/Z$, $\bar{g}_n=10.9/N$ are parametrized to reproduce the experimental odd-even mass differences
for nearby spherical nuclei in the number-projected BCS approximation~\cite{ALH08}.
The quadrupole, octupole and hexadecupole interaction terms have strengths given
by $\chi_\lambda= k_\lambda\chi$ for $\lambda=2,3,4$ respectively. The paprameter $\chi$ is determined self-consistently~\cite{ALH96} and $k_\lambda$ are
renormalization factors accounting for core polarization effects.
We use the parametrization of Ref.~\cite{OZE12} for $\gamma$ and $k_2$
 \begin{subequations}
 \begin{equation}
 \gamma = 0.72-\frac{0.5}{(N-90)^2+5.3}\;,
 \end{equation}
 \begin{equation}
    k_2 = 2.15+0.0025(N-87)^2 \;,
 \end{equation}
  \label{Eq:params}
 \end{subequations}
 while the other parameters are fixed at $k_3=1$ and $k_4=1$.

\section{Ground-state Energy}

Because of the sign problem introduced by the projection on an odd number of particles,
the thermal energy $E(\beta)$ of the odd-even samarium and neodymium isotopes can in practice be calculated only up to $\beta \equiv 1/T \sim 4-5$ MeV${}^{-1}$. For comparison, SMMC calculations in the neighboring even-even samarium and neodymium nuclei, for which there is no sign problem, were carried out up to $\beta=20$ MeV${}^{-1}$~\cite{OZE12}. Systematic errors
 introduced by the discretization of $\beta$~\cite{LANG93}, are corrected by calculating $E(\beta)$ for the two time slices of
$\Delta\beta=1/32$ MeV${}^{-1}$ and $\Delta\beta=1/64$ MeV${}^{-1}$ and then performing a linear extrapolation to $\Delta\beta=0$. For $\beta\gtrapprox 3$  MeV$^{-1}$, the dependence of $E(\beta)$ on $\Delta\beta$ is weaker and an average value was taken instead. The calculations for $\beta>2$ MeV$^{-1}$
were carried out using a stabilization method of the one-body canonical propagator~\cite{ALH08} for a given configuration of the auxiliary fields.

Since our calculations of the thermal energy $E(\beta)$ are limited to $\beta\sim 4-5$ MeV${}^{-1}$, we cannot obtained a reliable estimate of the ground-state energy $E_0$ in direct SMMC calculations. The Green's function method of Ref.~\cite{MUK12} was used successfully in medium-mass nuclei to circumvent the odd particle-number sign problem and extract accurate ground-state energies. However, this method becomes computationally intensive in heavy nuclei and requires additional development. Here we extract $E_0$ by performing a one-parameter fit of the SMMC thermal excitation energy $E_x(T)=E(T)-E_0$ to the experimental thermal energy. The latter is calculated using the thermodynamical relation $E_x(\beta)=-d\ln Z(\beta)/d\beta$, where $Z$ is the experimental partition function in which the energy is measured relative to the ground state (see below).

We demonstrate the method for $^{147}$Nd. Fig.~\ref{fig:ZE} shows the thermal excitation energy (left panel) and the partition function (right panel)
as a function of temperature. The SMMC results (open circles) are shown down to $T=0.25$ MeV, below which the statistical errors become too large due to the odd particle-number sign problem. We calculated the experimental partition
function (right panel, dashed line) from $Z(T)=\sum_{i} (2J_i+1) e^{-E_{x,i}/T}$ where $E_{x,i}$ denote the excitation energies of the experimentally known energy levels. The incompleteness of the level counting data above a certain excitation energy leads to a saturation of the experimental partition function and the corresponding average  experimental thermal energy (left panel, dashed line) above $T\sim 0.15$ MeV. In order to determine the ground-state energy $E_0$ from a fit of the SMMC thermal energy to the experimental data, it is necessary to have a realistic estimate of the experimental thermal energy above $T \sim 0.25$ MeV where SMMC results are available.
\begin{figure}[t!]
\begin{center}
%\vspace{0.6cm}
%\resizebox{0.90\columnwidth}{!}{
%  \includegraphics{E_Z.eps} }
  \includegraphics[clip,angle=0,width=1.0\columnwidth]{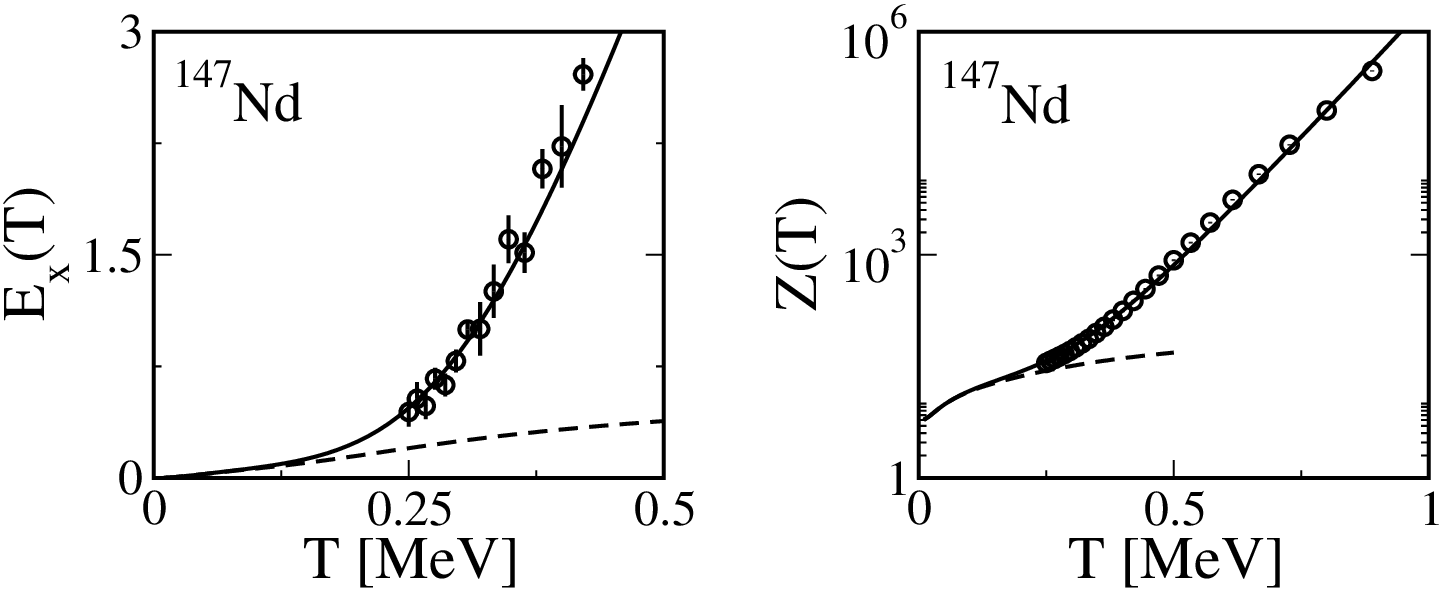}
\caption{Thermal excitation energy $E_x$ (left panel) and partition function $Z$ (right panel) versus temperature  $T$ for $^{147}$Nd. The SMMC results (open circles) are compared with the results deduced directly from experimentally known levels (dashed lines), and from Eq.~(\ref{Eq:ZBBF})
(solid lines). The ground-state energy $E_0$ is obtained by a one-parameter fit of the SMMC thermal energy to the solid line on the left panel.}
\label{fig:ZE}       % Give a unique label
\end{center}
\end{figure}
This can be accomplished by calculating an empirical partition function
\begin{equation}
   Z(T)=\sum_{i}^{N} (2J_i+1) e^{-E_{x,i}/T} +\int_{E_{N}}^\infty d E_x \,\rho_{\rm BBF}(E_x) e^{-E/T}\;,
   \label{Eq:ZBBF}
\end{equation}
where the discrete sum is carried out over a suitably chosen complete set of $N$ experimental levels up to an excitation energy  $E_N$, and the contribution of higher-lying levels is included effectively through the integral over an empirical level density with a Boltzmann weight. The empirical state density is parametrized by the back-shifted Bethe formula (BBF)~\cite{DILG73}
\begin{equation}\label{Eq:BBF}
\rho_{\rm BBF}(E)=\frac{\sqrt{\pi}}{12}a^{-1/4}(E-{\Delta})^{-5/4}e^{2\sqrt{a(E-{\Delta})}} \;,
\end{equation}
where $a$ is the single-particle level density parameter and $\Delta$ is the backshift
parameter. For these parameters we use the values in Ref.~\cite{OZE12-2} determined from level counting data
(at low excitation energies) and the $s$-wave neutron resonance data (at the neutron resonance threshold) for the given nucleus of interest. The solid lines in Fig.~\ref{fig:ZE} are calculated using Eq.~(\ref{Eq:ZBBF}) for the empirical partition function. The number of the low-lying states $N$ (determining $E_N$) are chosen such that the solid and dashed curves for the experimental partition function merge smoothly at  sufficiently low temperatures. The values of $a$ and $\Delta$ and the suitably chosen values of $N$  are tabulated in Table~\ref{table:params} for the nuclei of interest.
The ground-state energy $E_0$ is determined by
fitting the SMMC thermal excitation energy (open circles) to the solid curve in the left panel of Fig.~\ref{fig:ZE}. Table~\ref{table:params} also shows the fitted values of $E_0$ for the corresponding neodymium and samarium isotopes.

\begin{table*}[htb]%
\renewcommand{\tabcolsep}{0.084\columnwidth}
\renewcommand{\arraystretch}{1.2}
\begin{tabular}{@{}cccccc}
\hline
Nucleus & $a$ (MeV$^{-1}$) & $\Delta$ (MeV) & $N$ & $E_N$ (MeV) & $E_0$ (MeV) \\
\hline
${}^{143}$Nd &  15.951 & 0.759 & 3 & 1.228  & $-191.609 \pm 0.020$ \\
${}^{145}$Nd &  16.792 & 0.135 & 4 & 0.658  & $-210.010 \pm 0.028$ \\
${}^{147}$Nd &  18.276 & 0.058 & 4 & 0.190  & $-227.870 \pm 0.029$ \\
${}^{149}$Nd &  18.552 & -0.644 & 3 & 0.138 & $-245.440 \pm 0.021$ \\
${}^{149}$Sm &  19.086 & -0.196 & 8 & 0.558 & $-244.894 \pm 0.026$ \\
${}^{151}$Sm &  18.999 & -0.771 & 8 & 0.168 & $-264.968 \pm 0.020$ \\
${}^{153}$Sm &  17.848 & -1.053 & 7 & 0.098 & $-284.796 \pm 0.020$ \\
${}^{155}$Sm &  17.067 & -0.834 & 7 & 0.221 & $-304.245 \pm 0.025$ \\

\hline
\end{tabular}%
\caption{ The values of $a$ and $\Delta$ in the BBF [see Eq.~(\ref{Eq:BBF})] as determined from the level counting data
  at low excitation energies and the neutron resonance data~\cite{OZE12-2}, the number $N$ of a complete set of experimentally known levels
  and its corresponding energy $E_N$ used in Eq.~(\ref{Eq:ZBBF}), and the extracted ground-state energies $E_0$ for the odd-mass $^{143-149}$Nd
  and $^{149-155}$Sm isotopes.\label{table:params}}
\end{table*}%

We emphasize that only a single parameter $E_0$ is adjusted to fit the empirical partition function. Nevertheless, we see that both the SMMC partition function and thermal energy fit very well the corresponding experimental quantities over a range of temperatures. This is an evidence that our SMMC results are consistent with the experimental data.

\begin{figure}[t]
\begin{center}
%\vspace{0.6cm}
%\resizebox{0.8\columnwidth}{!}{
%  \includegraphics{Nd-all-rho.eps} }
\includegraphics[clip,angle=0,width=1.0\columnwidth]{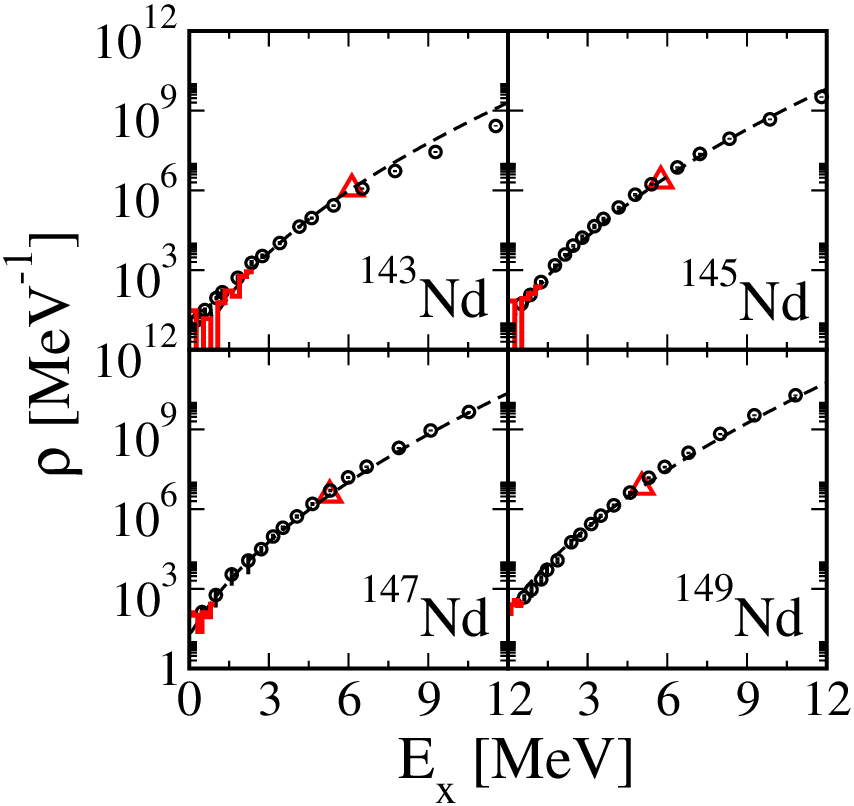}
\caption{State densities of odd-mass $^{143-149}$Nd isotopes versus excitation energy $E_x$. The SMMC densities (open circles) are compared with experimental results. The histograms at low excitation energies are from level counting data, the triangle is the neutron resonance data and the dashed line is the BBF state density [Eq.~(\ref{Eq:BBF})] that is extracted from the experimental data.}
\label{fig:rho-Nd}       % Give a unique label
\end{center}
\end{figure}

\section{State Densities}

\begin{figure}[tbh!]
\begin{center}
%\vspace{-0.2cm}
%\resizebox{0.8\columnwidth}{!}{
%  \includegraphics{Sm-all-rho.eps} }
\includegraphics[clip,angle=0,width=1.0\columnwidth]{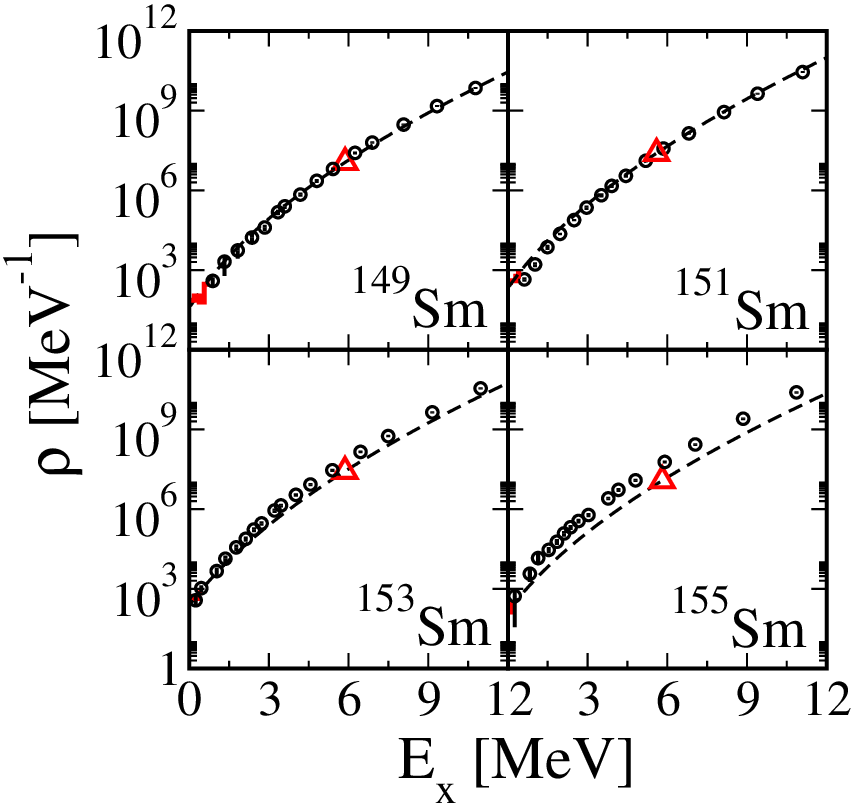}
\caption{State densities of odd-mass $^{149-155}$Sm isotopes versus excitation energy $E_x$. Symbols and lines as in Fig.~\ref{fig:rho-Nd}.}
\label{fig:rho-Sm}       % Give a unique label
\end{center}
\end{figure}

The average state density is determined in the saddle-point approximation to the integral that expresses the state density as an inverse Laplace transform of the partition function. This average density is given by
\begin{equation}
    \label{eq:levdens}
    \rho(E) \approx \frac{1}{\sqrt{2 \pi T^2 C}} e^{S(E)} \;,
\end{equation}
where  $S(E)$ is the entropy and $C$ is the heat capacity in the canonical ensemble.
In SMMC we first calculate the thermal energy as an observable $E(\beta)=\langle H \rangle$ and the integrate the thermodynamic relation
$-d\mathrm{ln} Z/d\beta = E(\beta)$ to find the partition function $Z(\beta)$. The entropy and
the heat capacity are then calculated from
\begin{equation}
S(E)=\mathrm{ln}Z+\beta E\;;\;\;\;\;\; C=-\beta^2 \partial E/\partial\beta
\end{equation}
and substituted into Eq.~(\ref{eq:levdens}) to yield the state density.
The SMMC state densities for the odd-mass neodymium and samarium isotopes are shown by open circles in Fig.~\ref{fig:rho-Nd} and Fig.~\ref{fig:rho-Sm}, respectively.
We compare these SMMC densities with the level counting data (histograms)
and with the neutron resonance data (triangles). We also show the empirical BBF level densities (dashed lines).
For the neodymium nuclei (Fig.~\ref{fig:rho-Nd}), we find excellent agreement among the SMMC results (open circles), the experimental state densities (both at the level counting and at the neutron resonance energies) and the BBF state densities (dashed lines).
In the case of the samarium nuclei (Fig.~\ref{fig:rho-Sm}), a similarly good agreement is observed for $^{149}$Sm and $^{151}$Sm. We also find overall good agreement for  $^{153}$Sm and $^{155}$Sm, although we observe some discrepancies between the
SMMC and the experimental state densities at the neutron resonance energy. Since the neutron resonance data point is used to determine
the parameters of the BBF state densities, similar discrepancies are observed between the SMMC and the BBF state densities.

\section{Conclusion}

We have extended the SMMC calculations in the rare-earth region to include the odd-mass isotopic chains of $^{149-155}$Sm and $^{143-149}$Nd. The sign problem introduced by the projection on an odd number of particles makes it difficult to calculate directly the ground-state energy. We circumvent this problem in practice by carrying out a one-parameter fit of the SMMC thermal excitation energy to the experimental average thermal excitation energy as determined from the neutron resonance data and level counting data at low excitation energies. We then calculate the state densities of the even-odd samarium and neodymium isotopes and find them to be in good overall agreement with state densities extracted from available experimental data. This method requires sufficient level counting data at low excitation energies and the neutron resonance data point at the neutron resonance threshold. The good overall agreement between the SMMC and experimental densities is an evidence that the theory is consistent with the data.

This work is supported in part by the U.S. DOE grant
No. DE-FG-0291-ER-40608 and as Grant-in-Aid for Scientific Research (C) No. 25400245 by the JSPS, Japan.  Computational cycles were provided by the NERSC high performance computing facility at LBL and by the facilities of the Yale University Faculty of Arts and Sciences High Performance Computing Center.

\end{document}